\documentstyle[sprocl,psfig]{article}

\def\Journal#1#2#3#4{{#1} {\bf #2}, #3 (#4)}

\def\PRL{\em Phys. Rev. Lett.}

\def\apj{{\em Astrophys. J.}}

\def\aap{{\em Astron. Astrophys.}}
\def\pasp{{\em Pubs. Ast. Soc. Pacific}}

\def\nat{{\em Nature}} 
\def\araa{{\em Ann. Rev. Astron. Astrophys.}} 

\def\be{\begin{equation}}
\def\ee{\end{equation}}
\def\bea{\begin{eqnarray}}
\def\eea{\end{eqnarray}}

\def\simlt{\mathrel{\hbox{\rlap{\hbox{\lower4pt\hbox{$\sim$}}}\hbox{$<$}}}}
\def\simgt{\mathrel{\hbox{\rlap{\hbox{\lower4pt\hbox{$\sim$}}}\hbox{$>$}}}}
\def\that{\,\widehat{t}\,}
\def\Msun{\, {\rm M_\odot}} 
\def\tmax{ t_{\rm max}} 
\def\umin{ u_{\rm min}} 
\def\Amax{ A_{\rm max}} 
\def\kpc {\, {\rm kpc}\,}
\def\kms {\, {\rm km \, s^{-1}}} 
\def\ten#1{\times 10^{#1}} 
\def\etal {{\em et al.}} 
\def\eff {{\cal E}} 

\begin{document}

\title{GRAVITATIONAL MICROLENSING RESULTS FROM MACHO}

\author{ W. SUTHERLAND }

\address{Dept. of Physics, 1 Keble Rd, Oxford OX1 3RH, UK} 

\author{ C. Alcock, R.A. Allsman, D. Alves, T.S. Axelrod, A.C. Becker, 
       D.P. Bennett, \\
   K.H. Cook, K.C. Freeman, K. Griest, J. Guern, 
   M.J. Lehner, S.L. Marshall,   \\ 
 B.A. Peterson, M.R. Pratt, P.J. Quinn, A.W. Rodgers, C.W. Stubbs, 
  D.L. Welch }

\author{   (The MACHO Collaboration)  } 
\address{ }


\maketitle\abstracts{
The MACHO project is searching for dark matter in the form
 of massive compact halo objects (Machos), by monitoring
 the brightness of millions of stars in the Magellanic Clouds
 to search for gravitational microlensing events. 
 Analysis of our first 2.3 years of data for 8.5 million stars in 
 the LMC yields 8 candidate microlensing events, 
 well in excess of the $\approx 1$ event expected from lensing by 
 known low-mass stars. The event timescales range from 34 to 145 days, 
 and the estimated optical depth is $\sim 2 \ten{-7}$, about half
 of that expected from a `standard' halo. 
 Likelihood analysis indicates the typical lens mass is 
 $0.5^{+0.3}_{-0.2} \Msun$, suggesting they may be
 old white dwarfs. 
 }

\section{Introduction} 

Since this is the first microlensing talk at a meeting 
comprising a majority of particle physicists, we will provide
a short introduction to microlensing before describing 
the main results from the MACHO project. 

The field of microlensing has undergone a dramatic expansion 
in the few years since the first candidates were discovered, 
and there are now a large number of results towards both the 
LMC and the Galactic bulge; detailed reviews are provided by 
refs.~\cite{pac-rev,rm-rev}. 
Updated information on the MACHO project, 
and links to other microlensing projects 
are available on our WWW site~\cite{macho-www}. 


In \S2 we discuss the motivation for Macho searches, in \S3 we outline
the basics of microlensing, and in the remainder of the paper we 
summarise the MACHO project, focusing on recent 
results from the 2-year LMC data. 

\section{Motivation}

While the most popular theories of galaxy formation involve 
a universe dominated by non-baryonic dark matter, 
we should keep in mind that there are really two dark matter problems, 
as emphasised by e.g. Turner~\cite{turner-pw}; 
the `first' dark matter problem is that 
baryon density inferred from primordial nucleosynthesis 
$\Omega_B \sim 0.02 - 0.08$ is
greater than that observed in 
stars and gas $\Omega_{\rm vis} \simlt 0.01$. 
The `second' dark matter problem is that the matter density
inferred from galaxy clusters and 
large-scale streaming motions is $\Omega_{\rm dyn} \simgt 0.2$ 
which is greater than $\Omega_B$.  
This suggests that the universe contains both baryonic and 
non-baryonic dark matter. 

If large numbers of unseen baryons exist, the most natural place
for them to hide is as compact objects in the halos of galaxies; 
hence the generic name `massive compact halo objects' or Machos. 
A wide variety~\cite{carr} of Macho candidates have been proposed; 
these include `brown dwarfs' which are
balls of H and He below the minimum mass $\sim 0.08 \Msun$
for fusion to occur; stellar remnants such as white
dwarfs or neutron stars; and black holes, which may be either
primordial or remnants. 
Even if Machos are abundant, they would be very hard to 
detect directly; some types of Macho such as brown dwarfs 
at $\simgt 0.01 \Msun$ or old white dwarfs may soon be constrained
by deep surveys in near-IR wavebands, 
but Jupiter-mass brown dwarfs or black holes would be almost
impossible to detect directly. 

In a classic paper, Paczynski~\cite{pac86} proposed that Machos
could be detected by their gravitational `microlensing' influence
on the light from distant stars; this led directly to 
the first generation of 
microlensing searches (EROS-1, MACHO and OGLE) which started observations
in the early 1990's and turned up the first microlensing 
candidates in 1993. 
More recently, several new projects are underway, 
including DUO, EROS-2~\cite{rich}, 
MOA, AGAPE and Vatt-Columbia.

\section{Microlensing} 

The principle of microlensing is simple; if a compact object lies
near the line of sight to a background star, the well-known GR light
deflection occurs, and two images of the star are formed on opposite
sides of the lens.  
For galactic scales, the angular splitting of these images
is $\sim 0.001$ arcsec which cannot be resolved at present (hence 
`microlensing'); 
but the two unresolved
images combine to increase the apparent brightness of the source. 
The characteristic length-scale is the `Einstein radius' 
\be
 r_E = \sqrt{4 G m L x (1-x) \over c^2} 
\ee
where $m$ is the lens mass, $L$ is the source distance 
and $x$ is the ratio of the lens and source distances. 
For a source in the LMC at $L = 50 \kpc$ and a lens at $10 \kpc$, 
$r_E \approx 10^9 \, {\rm km}\, \sqrt{m / \Msun}$;   
since this is much larger than a typical star or Macho, in most cases 
we may assume a point source and point lens, and the resulting
magnification is simply~\cite{refsdal}  
\be 
A = { u^2 + 2 \over u \sqrt{u^2+4} } 
\ee
where $ u = b / r_E$ and $b$ is the distance of the lens from the
undeflected line of sight. 
For $u \simlt 0.5$, $A \approx u^{-1}$, 
so the magnification may be large; while it drops rapidly 
as $A \approx 1 + 2 u^{-4}$ for $u \gg 1$. 

Of course, a constant magnification is not detectable since
we don't know the intrinsic brightness of the source; but 
due to the relative motion of observer, lens and source, 
the magnification is transient with a duration 
\be
\that \equiv 2 r_E / v_\perp \sim 130 \,\sqrt{m \over \Msun} \ {\rm days},
\ee
where $v_\perp \sim 200 \kms$ is the transverse velocity of the lens
relative to the line of sight. 
This is a convenient timescale for astronomical observations. 
Also, the dependence $\propto \sqrt{m}$
means that by monitoring on a range of timescales, the experiment
may be sensitive to a wide range of masses from $\sim 10^{-7} \Msun$
to $\sim 100 \Msun$, covering most of the popular Macho candidates. 
This mass range is set at the low end 
where $r_E$ is smaller than the size
of a typical star and large magnifications cannot occur; and at the high 
end where the event duration exceeds the few-year duration
of a typical experiment. 
(Other lensing techniques are sensitive to 
different mass ranges; e.g. VLBI searches for macrolensed quasars, 
and searches for microlensing of quasars~\cite{schild} 
or perhaps gamma-ray bursts). 

\subsection{Optical Depth} 

Since $r_E \propto \sqrt{m}$, the solid angle subtended by a lens
at a given distance is $\propto m$; thus, the probability that 
a random star is microlensed with $u < 1$ or $A > 1.34$ at any instant
depends on the mass density of lenses $\rho(l)$ along the line of sight, 
but not their individual masses. 
This probability is called the `optical depth' $\tau$, and
is given by 
\be 
\tau = {4 \pi G \over c^2} \int_0^L \rho(l) {l (L-l) \over L} dl 
\ee
By the virial theorem, 
it is easily shown that $\tau \sim v^2 / c^2$ where $v$ is the
orbital velocity of the Galaxy. More detailed calculations
\cite{griest91} give an optical depth 
towards the Large Magellanic Cloud of
\be
\tau_{\rm LMC} \approx 5 \times 10^{-7}
\ee
for an all-Macho halo of `standard' form. 
(This number is uncertain by perhaps $50\%$ due to uncertainties
in the halo model.  However, as a reasonable 
approximation it scales proportional to the halo mass inside $50 \kpc$; 
it is not too sensitive to the halo flattening or core radius). 
Note, however, that the event rate $\Gamma$ does depend on the lens
masses $\propto m^{-0.5}$, because 
\be 
\Gamma = 4 \tau / \pi \langle \that \rangle \approx 
   1.6\ten{-6} \, (m/\Msun)^{-0.5} {\rm \, events/star/year} 
\label{eq-rate} 
\ee
i.e., low-mass Machos produce (relatively !) numerous short events
whereas massive Machos produce fewer long-lasting events. 

The very low optical depth is the main difficulty of the experiment; 
only one star in two million will be magnified by $A > 1.34$ at any
given time, while the fraction of intrinsic variable stars is 
much higher, $\sim 0.3 \%$. 
Fortunately, microlensing events have many strong signatures
which are different from all currently known types of stellar 
variability. Assuming a single point source and lens, and uniform motions, 
the events should have a symmetrical shape 
given by eq. (2) and 
$u(t) = [ \umin^2 + ((t - \tmax)/0.5\that)^2]^{0.5}$, 
they should be achromatic, and at most one event should
be seen in any given star since the probability is so low. 

In reality, various deviations may occur due to e.g. 
blending of the source star with other unresolved stars,  
a binary lens or source, 
the non-uniform motion of the Earth, or the finite size
of the source; but the above form should be a good approximation for
most events.  

If many events are found, several statistical checks can also be made;
allowing for the detection efficiency,  
the events should be randomly distributed across
the colour-magnitude diagram, 
the distribution of peak magnifications should correspond to 
a uniform distribution in $\umin$, 
and the event timescales and peak magnifications should be uncorrelated. 

\section{Observations} 

Due to the low optical depth, a very large number of stars must
be monitored over a long period to obtain significant results. 
The simplest targets for this search are the Large and Small 
Magellanic Clouds, the largest of the Milky Way's satellite galaxies, 
since they have a high surface density of stars, they are distant enough
at 50 and 60 kpc to provide a good path length through the dark halo, 
and they are located $30^o$ and $45^o$ from our galactic plane, so
the density along the line of sight is dominated by dark matter.  
We also observe the Galactic Bulge when the LMC and SMC are too low in
the sky. 

Since mid-1992, the MACHO collaboration has had full-time use of the 
1.27-m telescope at Mt.~Stromlo Observatory near Canberra, Australia; 
an extended run until 1999 has recently been approved. 
Details of the telescope are given by ref.~\cite{macho-tel}, and of the
camera system by ref.~\cite{macho-marshall}. 
Briefly, an optical corrector gives
a field of view of $0.7 \times 0.7$ degrees, and a dichroic 
beamsplitter is used to take simultaneous images in red and blue passbands. 
The two foci are equipped with very large 
CCD cameras, each containing 4 CCD chips of 
$2048 \times 2048$ pixels. 
The typical exposure time is 300 sec, and about 60 images are taken 
per clear night; 
we took our 50,000th image in October 1996. 
All the 4~TB of raw data is archived to Exabyte tape. 

A special-purpose code~\cite{macho-lmc1} is used to measure the brightness of 
all stars in the images; briefly, one good-quality image of each
field is used to define a `template' list of stars. Each subsequent
image is aligned with the template using bright reference stars, 
and a point spread function (PSF) is estimated from these. 
Then, the flux of all stars is estimated using the known
positions and PSF; this provides a dramatic time saving 
as well as more accurate results. The reductions take around
1 hour per image on a Sparc-10. 

Since late 1994 we have implemented same-day 
processing for a large fraction of our fields, 
which enables us to detect events in 
real time~\cite{macho-pratt,alert-www}. 

Just to mention our Galactic Bulge results: 
we have detected over 100 microlensing events towards the
bulge~\cite{macho-bulge2,alert-www}, including several 
events due to binary lenses  
and one showing asymmetry due to the Earth's orbit~\cite{macho-parallax}. 
Although the lensing towards the bulge 
is probably dominated by low-mass stars rather than dark matter, 
this has interesting consequences for Galactic structure, 
as well as providing a very nice proof of microlensing. 

\section{LMC Results} 

We have recently completed an analysis~\cite{macho-lmc2}
of the first 2.3 years of data for
22 well sampled LMC fields; this comprises over 8 million stars
with 300 to 800 observations each. 
We select microlensing candidates using a set of objective
selection criteria. 
The most important of these are that the star should have a
brightening of high significance, with peak magnification 
$\Amax > 1.75$, and that its flux should be 
approximately constant outside this region. 
[These selection criteria have been modified 
since the first year's LMC analysis~\cite{macho-lmc1}, 
due in part to experience with the bulge events. 
Briefly,  we have relaxed the cuts on the `standard' microlensing 
 shape, achromaticity and stellar crowding, 
 but we now require higher significance and magnification; 
thus, the `marginal' events 2 \& 3 from refs.~\cite{macho-prl,macho-lmc1}
do not pass the new cuts, but some new first-year events appear.]

\begin{figure}
\psfig{figure=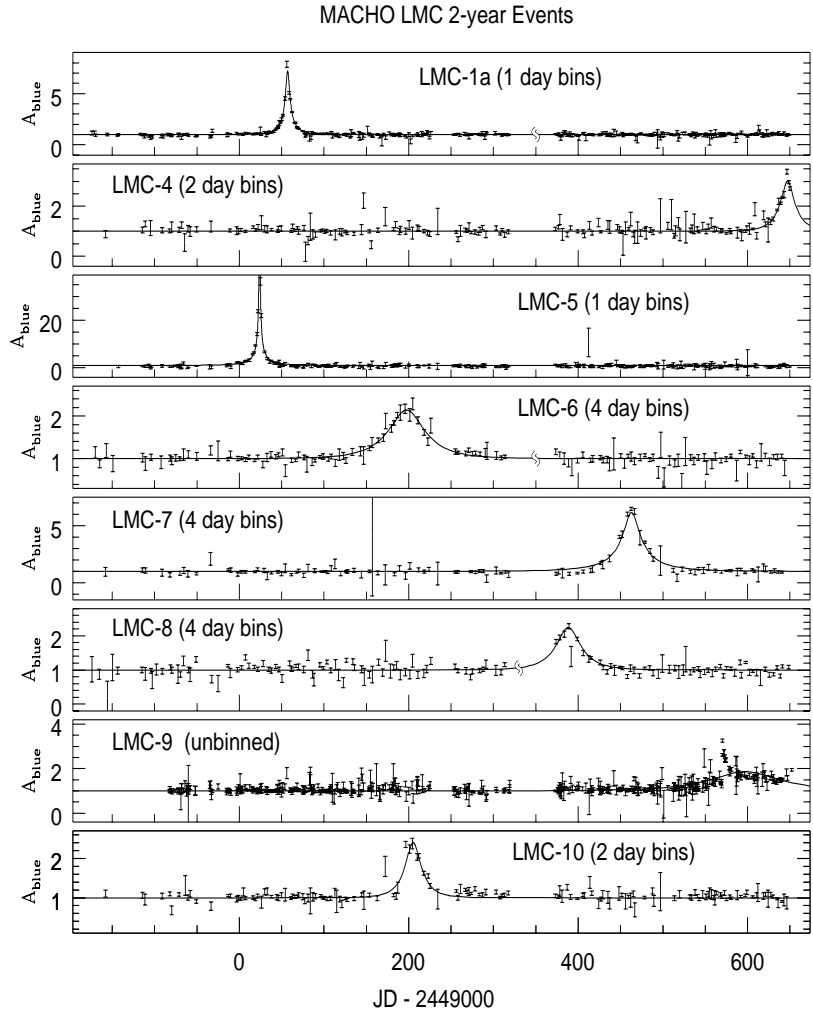,height=5.5in,width=4.6in}
\caption{Lightcurves of the 8 candidate events from the 2-year LMC data. 
 Flux measurements (one colour) are plotted in linear 
 units with 1$\sigma$ errors, averaged in time bins (see labels) for clarity, 
 and normalised to the fit baseline for each star. 
 The curves show the single-lens microlensing fit.  Time is in days. 
\label{fig-events} }
\end{figure}

We find 12 objects in the 2-year dataset passing the final cuts, 
of which 4 correspond to 2 stars doubly detected in field overlaps, 
and 2 are rejected due to `magnification bias' in that they were 
brighter than normal in the template image and then faded below 
our detection limit (one of these was superposed on 
a background galaxy, and was almost certainly a supernova). 
Thus we have 8 microlensing candidates, with timescales
from 34 to 145 days, shown in Figure~\ref{fig-events}; 
they are numbered 1, 4-10 to avoid ambiguity. 
Of these eight candidates, six are well 
fitted by the standard microlensing 
shape; three of them (numbers 5, 7 and 9) 
show evidence of chromaticity, but this is found to be consistent 
with blending~\cite{macho-lmc2}. 
Event~9 shows a distinctive double-peaked structure
 and is clearly~\cite{bennett-sm} due to a binary lens.
\footnote{A binary lens can produce a great diversity of possible
lightcurves~\cite{mao-pac}. 
However, `caustic crossings' are generic features; these
occur where the number of images
changes from 3 to 5 or vice versa, and the magnification becomes
large when the source is just inside the caustic. Since the caustic(s)
are closed curves in the source plane, 
caustic crossings must occur in inward/outward pairs.}
Event~10 is somewhat asymmetrical and 
may be a variable star, though it could also be microlensing 
of a binary source star.
The inclusion or exclusion of this event has
little influence on the results.

We are confident that most of our 8 candidates are genuine
microlensing events; they cannot be due to observational error, 
cosmic ray hits, satellite trails etc. 
since they are seen at different pixel locations (due to pointing
variations) in dozens of independent CCD frames; 
event~1 was confirmed by EROS, and event~4 was detected in real-time
and observed with other telescopes. 
 
Intrinsic stellar variability is more difficult to exclude, but
several of the candidates have high magnifications, event~9
is very characteristic of a binary lens, and event~4 
was observed spectroscopically and appeared normal. 
The distribution of peak magnifications and in the colour-magnitude
diagram is consistent with expectation~\cite{macho-lmc2}. 
This test also suggests that at 
least 5 of the candidates are genuine microlensing, 
since if {\it only} the `high-quality' candidates (e.g. 1,4,5 and~9) 
were microlensing, the distribution of $\Amax$ would 
be somewhat improbable~\cite{macho-lmc2}.  

\section{Implications} 

\begin{figure}
\centerline{\hbox{ 
\psfig{figure=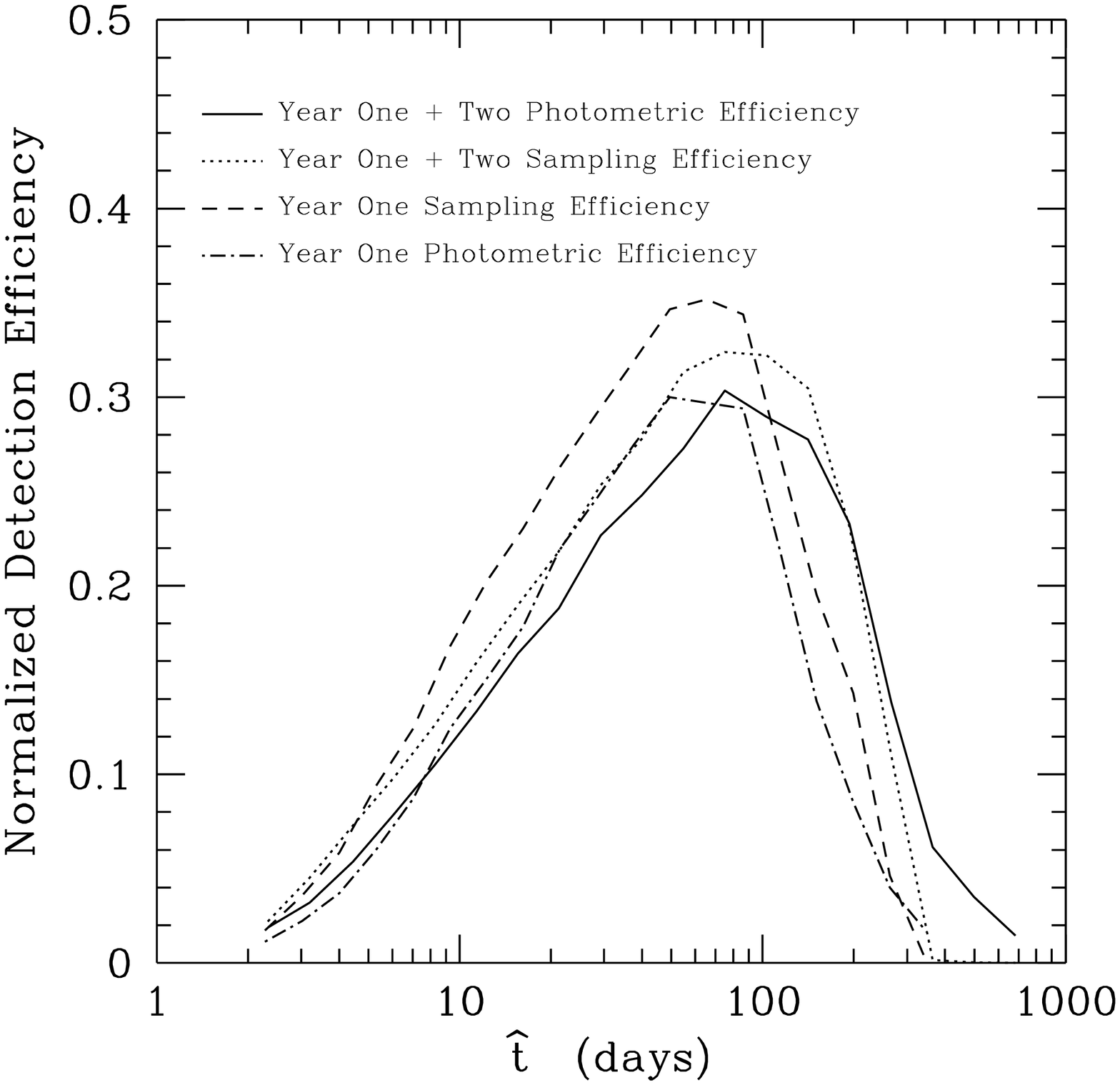,height=2in}
}}   
\caption{ The solid line shows the microlensing 
 detection efficiency (relative to event rate with $\umin < 1$) 
 for LMC 2-year sample. 
 \label{fig-eff} }
\end{figure}

In order to derive quantitative results, we clearly need to 
know our detection efficiency.  
We have evaluated this using  a series of 
Monte-Carlo simulations; these include the addition of
 artificial stars at a range of magnifications into real data 
frames, and also incorporate the known times of observations
incorporating bad weather, variable seeing conditions etc. 
Simulated microlensing events are then processed through our
standard software to give the detection efficiency as a function
of the event timescale, $\eff(\that)$, shown in Figure~\ref{fig-eff}. 

It is convenient to show the expected number of events
assuming that all the halo is made of Machos with a unique mass $m$;  
this is given for a standard halo model in Figure~\ref{fig-lim}a. 
There are two competing effects: for larger masses $m \simgt 0.01 \Msun$,
most events have timescales $\that \simgt 10$ days where our efficiency is
quite good, but the event rate is falling $\propto m^{-0.5}$. 
For small masses $m < 0.001 \Msun$,  
the theoretical event rate is high but most events are shorter 
than $\that \sim 3$ days where our efficiency is low. 
The product of these two effects gives rise to the peak at
$\approx 45$ expected events for $m \sim 2 \ten{-3} \Msun$.

\subsection{Limits on Low-Mass Machos} 

Although the efficiency is falling towards short event durations, 
the absence of short events is still very significant, because
of the $m^{-0.5}$ factor in eq.~\ref{eq-rate}. 
From the fact that we have no candidate event with $\that < 20$ days
in the above data, we can conclude that Machos with masses
from $6 \ten{-5}$ to $0.02 \Msun$ contribute less than $20\%$ of the standard
halo at 95\% confidence. 
We have extended these limits to lower masses using a separate 
`spike' search~\cite{macho-spike} for very short-timescale events.
Some of our fields are observed twice per night, 
giving a set of 4 data points, two in each passband. 
We then search for events where all 4 data points
on such a night exceed some threshold, while no such deviation
occurs in the rest of the light curve. 
After suitable cuts, we find no such events, and this sets interesting
limits on events with durations $\sim 0.3 - 3$ days. 

\begin{figure}[htb]
\centerline{\hbox{ 
\psfig{figure=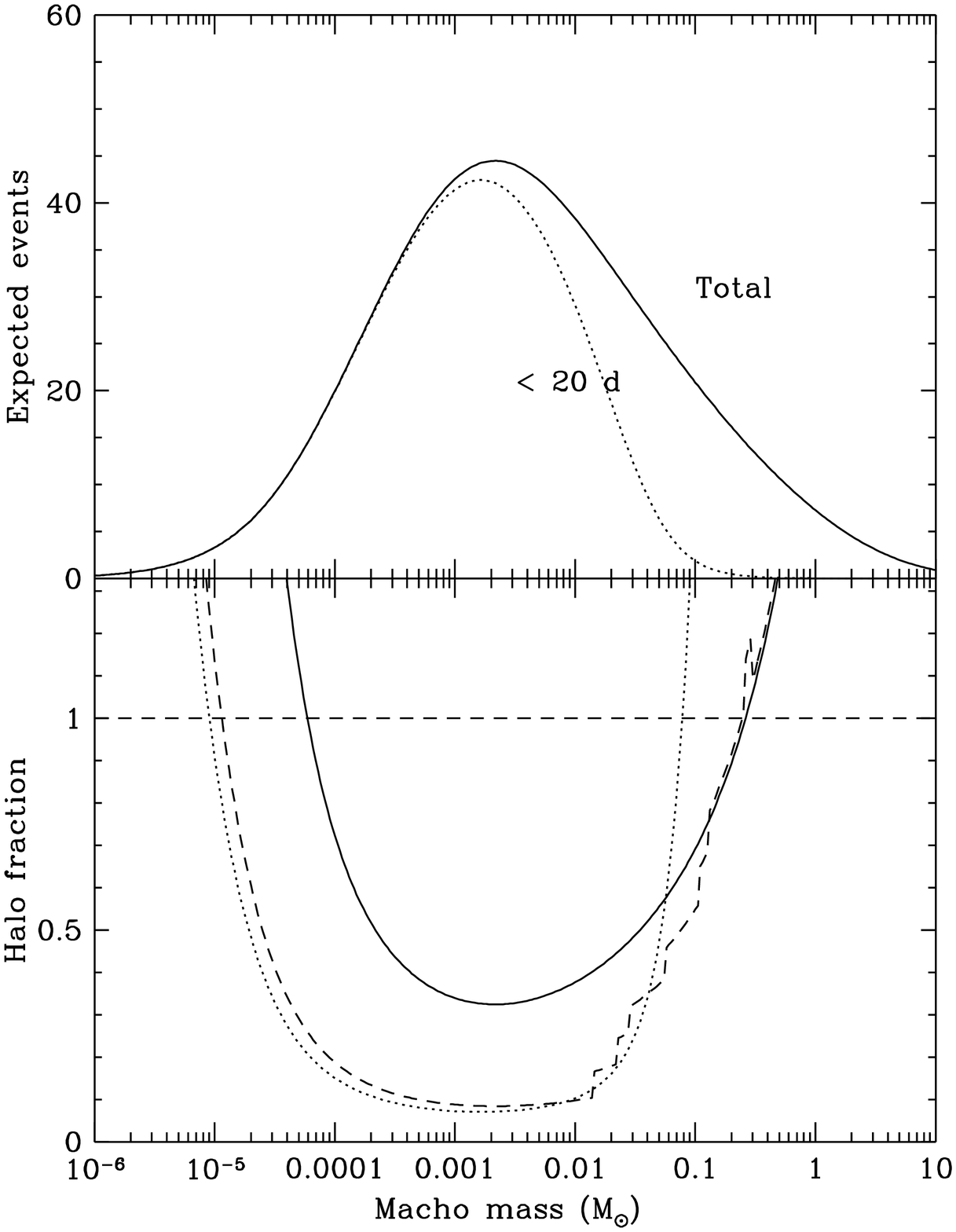,height=3in}
\psfig{figure=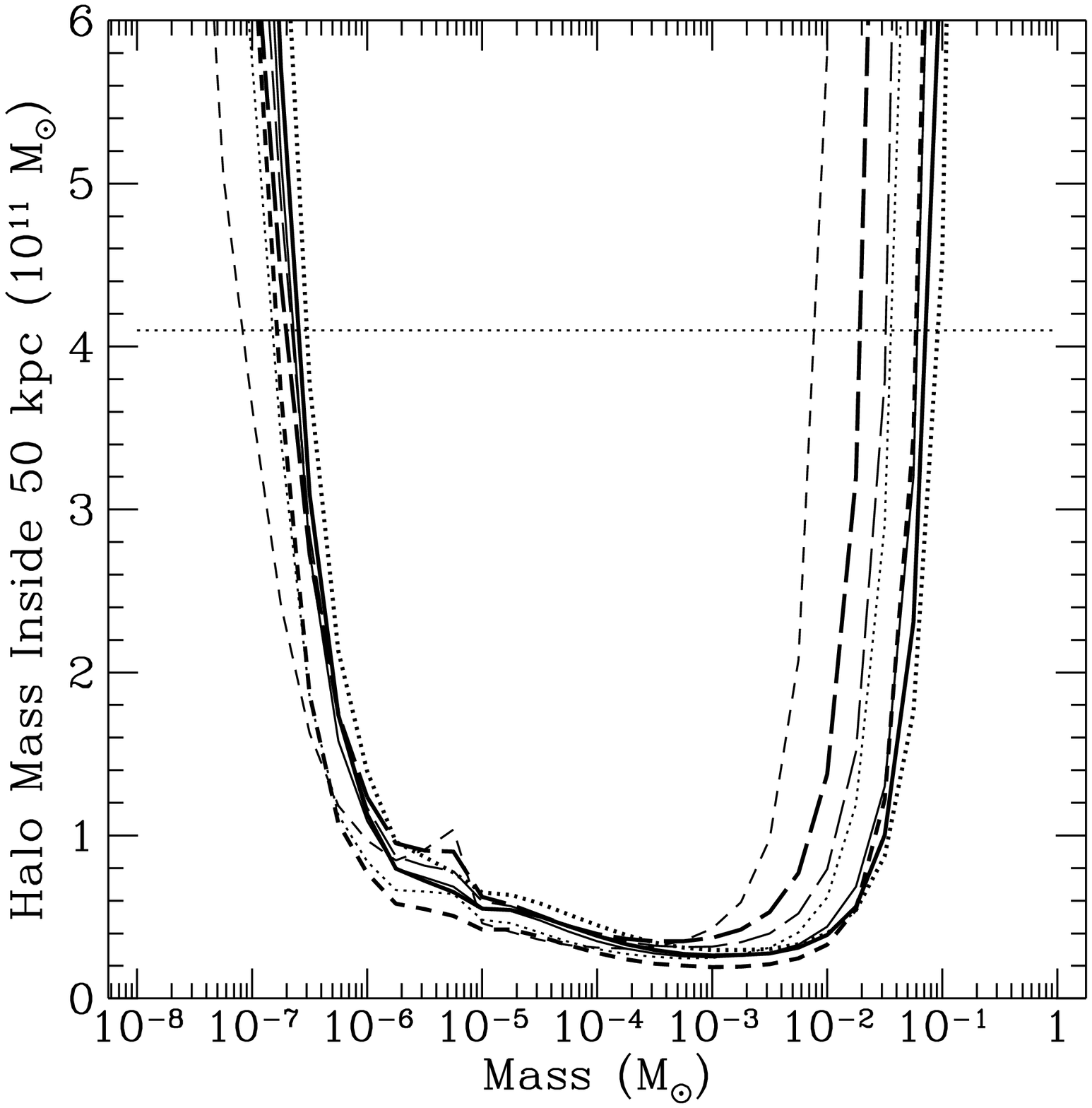,height=2.4in} 
}}   
\caption{ (a) Upper panel shows 
 expected number of events for all-Macho halo with unique Macho
 mass $m$. Lower panel shows derived limits on halo Macho fraction. 
 Regions above the curves are excluded at 95\% c.l. 
 The solid line is from 8 observed events, the dotted line from 
 no events with $\that < 20$ days.   \hfill 
 (b) Upper limits (95\% c.l.) on total mass 
 of Machos interior to $50 \kpc$, 
 from combined `spike' and standard analysis, for 8 halo
 models. 
\label{fig-lim} }
\end{figure}

Combining these two analyses, we conclude that Machos in the 
mass range $10^{-6}$ to $0.02 \Msun$ comprise 
less than 20\% of the standard halo; more generally, such objects contribute 
less than $10^{11} \Msun$ to the halo mass
within $50 \kpc$, as shown in Figure~\ref{fig-lim}b. 
Similar (nearly independent) limits have been derived by EROS~\cite{eros-lim}.

\subsection{Implications of the 8 Events}  

We can estimate the optical depth via 
 $\tau_{\rm est} = (\pi/4 E) \sum_i \that_i / \eff(\that_i) $, where
$E = 1.8 \ten{7}$ star-years is our `exposure', and 
$\that_i$ is the timescale of the $i$-th event. 
Accounting for our detection efficiency, 
the 8 events give an estimated optical depth 
of $\tau_{\rm est} = 2.9^{+1.4}_{-0.9} \ten{-7}$, which 
is just over half of that from an all-Macho dark halo.\footnote{
Although the event rate is similar to our earlier estimate
(3 candidates in the first year), 
the new optical depth estimate is considerably higher 
because events are `weighted' proportional to their duration, 
and the new events all have longer timescales.} 

For an event with the `standard' shape, it is not possible to 
tell where the lens is situated along the line of sight; thus, lensing
events can also arise from faint stars in our Galaxy and the 
LMC~\cite{sahu} itself, as well as halo Machos. 
However, lensing by known stars 
is expected to contribute only $1.1$ events in this sample, or 
$\tau_{\rm stars} \sim 0.5 \ten{-7}$, so there appears to be a very 
significant excess~\cite{macho-lmc2}. 
A more conservative estimate of the halo optical depth 
is given by excluding event~9
(since the lens may be in the LMC~\cite{bennett-sm}), and 
event~10 which may be a variable star; this 
gives $\tau_{\rm halo} = 2.1^{+1.1}_{-0.7} \ten{-7}$. 

\begin{figure}[htb]
\centerline{\hbox{ 
\psfig{figure=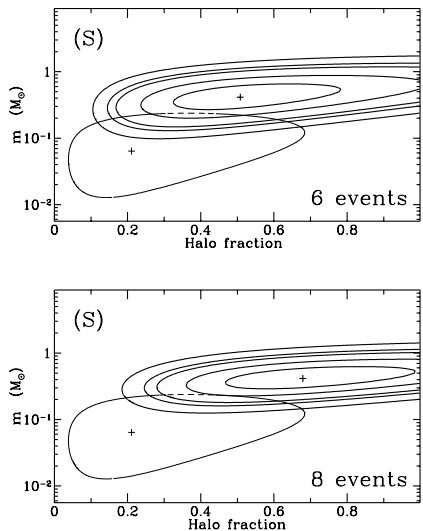,bbllx=50pt,bblly=50pt,bburx=550pt,bbury=800pt,height=3in}
}}   
\caption{ Bold lines are likelihood contours (34,68,90,95,99\%  
 enclosed probability) for Macho mass and Macho fraction 
 of the halo, for the standard halo model, for the 8 and 6 event
 samples. The light line shows the $90\%$ contour from the 1-year 
 analysis. 
\label{fig-like} }
\end{figure}

We can estimate the lens masses using the event
durations; since one observable $\that$ depends on three unknowns, the
lens mass, distance and transverse velocity, this is 
only a statistical estimate and is somewhat sensitive to the assumed 
halo model. For the standard halo, 
likelihood analysis (Figure~\ref{fig-like}) gives
a most probable lens mass of $0.5^{+0.3}_{-0.2} \Msun$. 
If the lenses are in the halo in this mass range, 
they cannot be hydrogen-burning stars which would be 
easily detectable~\cite{fgb}. 
Thus, remnants such as old white dwarfs appear to 
be a natural possibility. 
These are not excluded by star-count data, though they must be very faint. 
White dwarfs also require a rather narrow initial mass function in order
to avoid overproducing low-mass stars or supernovae, and may have 
problems with the high luminosities of the progenitor stars; 
thus, primordial black holes are a more exotic possibility. 

Although the formal significance of our number of candidates is high, 
we cannot yet claim a conclusive detection of dark matter; 
for instance, if a few 
of our lower-quality candidates were actually variable stars, 
and the stellar lensing rate were double our estimate above, 
we would still have an excess of events, 
but the significance would only be marginal. 
The lensing rate from stars in our own disk 
is directly constrained by HST star counts;
that from stars in the LMC is more uncertain, but is 
constrained by the line-of-sight
velocity dispersion~\cite{gould-self}, and improved measurements
are planned.  

Another loophole is that there might be a small dwarf galaxy 
located between us and the LMC~\cite{zhao}; this could account for
most of the observed optical depth, though the {\it a priori} 
probability of such an alignment is only $\sim 1\%$. 

There are a number of prospects for clarifying these results; 
we should soon have analysed another 2 years of data, 
and we will continue observations until 1999; this 
should give many more events, and also extend the
search to longer timescales. 
The real-time detection system also enables more precise
follow-up photometry and spectroscopy to check future candidates; 
two more real-time LMC events have been discovered during 1996. 
If most of the lenses reside in the LMC itself, then the
events should occur preferentially near the center of the LMC, 
while halo lensing will produce a more even distribution. 
Additionally, if the source star is a binary with a period
shorter than the event duration ($\sim 10\%$ of events), 
it is possible to test whether or not the lens 
is in the LMC~\cite{han-gould}.

If the lenses are old white dwarfs, they should be accessible
to deep searches using the HST or wide-field ground-based imaging. 
In the longer term, observations from a small satellite
in Solar orbit can measure the 
projected velocity of the lens~\cite{gould-1sat,gould-this}, 
or interferometric measurements could resolve the double images
and measure the angular Einstein radius; either one of these
measurements can determine whether the lens belongs to the
galactic disk, halo or the LMC, and both together would solve
for the lens mass, distance and velocity. 

In summary, we have found very interesting evidence
that Machos in the mass range $0.05 - 1 \Msun$ contribute a substantial
fraction of our Galaxy's dark matter; continued observations
should clarify this in the next few years. 

\section*{Acknowledgements}

This work has been supported at LLNL by DOE contract W7405-ENG-48, 
at the Center for Particle Astrophysics by NSF grant AST-8809616, 
and at MSSSO by the Australian Dept. of Industry, Technology and Regional
Development. WS is supported by a PPARC Advanced Fellowship.

\end{document}